**Annotated bibliography**

*Philosophy of Astrophysics: Stars, Simulations, and the Struggle to Determine What is Out There*
(Synthese Library)
eds. Siska De Baerdemaeker, Nora Boyd, Vera Matarese and Kevin Heng

Cameron C. Yetman
Department of Philosophy
University of Toronto

**Abstract**

The following annotated bibliography contains a reasonably complete survey of contemporary work in the philosophy of astrophysics. Spanning approximately forty years from the early 1980s to the present day, the bibliography should help researchers entering the field to acquaint themselves with its major texts, while providing an opportunity for philosophers already working on astrophysics to expand their knowledge base and engage with unfamiliar material.

**Author note**

https://orcid.org/0000-0002-4464-3337
https://philpeople.org/profiles/cameron-yetman

Correspondence concerning this chapter should be addressed to
cameron.yetman@mail.utoronto.ca.

**Introduction**

The bibliography is divided into seven sections. The first section covers methodological issues in astrophysics: how do astrophysicists make observations, interpret data, and solve problems which arise in the process? The section includes case studies on gravitational waves, astroparticle physics, dark matter, extra-galactic objects, and others.

The second and largest section covers topics related to astrophysical modelling and computer simulations: their epistemic value, their limits, and their application to major problems in the field. This section contains case studies on galactic modelling, analogue experiments, cosmological simulations, and code comparisons.

The third section concerns perhaps the oldest debate within the contemporary philosophy of astrophysics, namely between astrophysical realists and anti-realists, which was initiated by the work of Ian Hacking in the 1980s. The section contains case studies on astroparticle physics, gravitational lensing, dark matter, as well as stellar physics and classification.

The fourth section covers the relationships between astrophysical theory, observation, confirmation, and more. This section contains case studies on singularities, general relativity, dark matter, interstellar interlopers, and gravitational waves.

The fifth section is somewhat tangential to mainstream work in the philosophy of astrophysics, but nevertheless contains a number of articles of which philosophers should be aware and with which they should be prepared to engage. The section covers issues in the sociology of scientific knowledge (SSK), as well as other social issues related to astrophysics. The section contains case studies on gravitational waves, the Hubble Space Telescope, astronomy and high-energy physics, the Herschel Space Observatory, "star-crushing", the Gemini Telescopes, Pluto, and the use of visualizations.

The sixth section contains works on typicality, the anthropic principle, and extra-terrestrial life. Most of the existing literature on typicality has an explicitly cosmological focus, but philosophers of astrophysics may offer fresh perspectives to these debates. The articles were chosen due to their potential interest for philosophers in this field, though few have explicit astrophysical content. Section six, then, serves as an invitation for philosophers of astrophysics to explore a field largely untouched by those with their knowledge and skillset.

The seventh and final section, compiled by Siska De Baerdemaeker, explores recent work related to dark matter and MOND (Modified Newtonian Dynamics) on both astrophysical and cosmological scales. This section is by no means a comprehensive overview of the philosophical literature on MOND, but the entries included have been chosen for their specific relevance to philosophy of astrophysics.

At the end of each section, there is a list of articles which deal with the section's theme, but whose primary theme warranted placing them somewhere else.

The reader will notice a number of articles which focus on cosmology or astronomy, rather than astrophysics. These articles were chosen in virtue of their potential applicability to problems in the philosophy of astrophysics, as judged by myself in discussion with other philosophers in the field. Every effort was made to avoid inflating the bibliography beyond its natural bounds, but some material from adjacent fields was necessary to provide a comprehensive overview of the state and future of the philosophy of astrophysics.

Especially given the relatively small size and recent vintage of this field (there are only 79 entries in this bibliography, 58 of which are from 2010 onwards, and 27 since 2020) the articles in this volume constitute a significant and timely addition.

## 1. Methodologies in astrophysics

The first comprehensive study in the epistemology of gravitational wave (GW) astrophysics, Elder (Black Hole Initiative) discusses the distinction between "direct" and "indirect" observations of gravitational waves, raises a circularity problem facing model-dependent observations (and explains how it is mitigated by GW astronomers), and elaborates on the virtues of multi-messenger astrophysics for creating more robust dependency relations between sources and traces of data, among other topics.

The authors of the LIGO-Virgo collaboration's "discovery paper" for the binary black hole merger GW150914 claim to have made a "direct detection" of gravity waves and a "direct observation" of the merger. Elder (Black Hole Initiative) seeks to disambiguate the meaning of terms like "direct", "indirect", "observation" and "measurement" in a way which is both philosophically adequate and true to how scientists use these terms. Elder argues that the LIGO-Virgo team can only be said to have *indirectly* detected a binary black hole merger due to their reliance on model-based inferences, thereby raising some important epistemic challenges that gravitational wave astrophysicists must overcome.

Although cosmology proceeds top-down (from theory to data and from large-scale to small scale) and astroparticle physics proceeds bottom-up (from detection of particles to theorizing about their cosmic sources), Falkenburg (TU Dortmund) argues that these disciplines pursue complementary strategies of scientific explanation while aiming at theoretical unification – a fact inadequately captured by contemporary philosophical accounts of scientific explanation and realism. Given this, Falkenburg urges the philosophical community to pay greater attention to astroparticle physics and the way in which it contributes to the empirical basis of cosmology.

Groups studying WIMPs have generated apparently incompatible data. Hudson (University of Saskatchewan) argues that this data is only incompatible given certain ancillary assumptions involved in data processing, and that we can reconcile the discordant results into an empirically adequate model *à la* van Fraasen (we cannot be realists about this model).

Astrophysicists have argued on the basis of a "causal argument" that the rapid variability in the brightness of quasars requires that their sources be extremely compact. Salmon (d. 2001, form. University of Pittsburgh) identifies the "cΔt size criterion" – according to which the region of brightness-variation cannot be larger than the distance light travels in its time of variation – as a crucial premise in this causal argument. Salmon claims that scientists have treated this criterion (or at least have often appeared to treat it) as a law of nature derived from special relativity, but that in fact it is "egregiously fallacious". If the criterion has any use at all, it is as a plausibility principle for fixing Bayesian priors when attempting to construct quasar models, and *not* as a physical requirement which such models must satisfy.

Wilson (University of Melbourne) provides an overview of the missing satellites problem in galactic astrophysics and analyzes how researchers have attempted to solve the problem. According to Wilson, these researchers have "black-boxed" their simulations by treating them as self-contained worlds in which simulated phenomena are epistemically significant, and they have blended these simulated results with real-world observations in generating their solution to the problem. This process of blending can make simulated worlds not merely possible, but *plausible*.

*For further articles relevant to this category, see Boyd 2018, Curiel 2019, De Baerdemaeker and Boyd 2020, Gueguen 2020, Gueguen 2021, Massimi 2018, and Meskhidze 2017.*

## 2. Models and simulations

Anderl, S. (2018). Simplicity and simplification in astrophysical modeling. *Philosophy of Science, 85*(5), 819-831. https://doi.org/10.1086/699696.

> Should astrophysical models strive to be "complete" (i.e., to capture all the details of the available data), or simple? Anderl (*Frankfurter Allgemeine Zeitung;* Institut de Planétologie et d'Astrophysique de Grenoble) argues that in many cases, simplicity is a valuable representational ideal because simple models facilitate (1) faster, more comprehensive exploration of the parameter space, and (2) internal validation of a model and the concomitant use of "physical intuition" which is so important for good model building.

Bailer-Jones, D. M. (2000). Modelling extended extragalactic radio sources. *Studies in History and Philosophy of Modern Physics, 31*(1), 49-74. https://doi.org/10.1016/S1344-2198(99)00028-3.

> This article discusses practical and epistemological issues associated with scientific modeling of novel phenomena, using extended extragalactic radio sources (EERSs) as a case study. Bailer-Jones (d. 2006, form. University of Heidelberg) argues that models are ways of representing the causal mechanisms behind poorly understood phenomena ("representation [caus.mech.]"), and that they also serve as conventional means of representing the unity of explanations of such mechanisms ("representation [conv.]"). Although models are "a central form of knowledge about empirical phenomena" (69), they can rarely be taken to constitute a definitive statement of what the world is really like; their epistemological status is thus quite complicated.

Boyd, N. M. (2015). Are astrophysical models permanently underdetermined? [Unpublished manuscript]. http://jamesowenweatherall.com/wp-content/uploads/2014/10/Boyd_SoCal_060615.pdf.

Against Hacking (1989) and Ruphy (2011), Boyd (Siena College) argues that we ought to be more optimistic about the prospects of breaking underdetermination in representation-driven astrophysical modeling. Using case studies from research into supernovae, dark matter, structure formation, and gamma ray bursts, Boyd articulates a framework according to which models with identifiable distinguishing features can be evaluated separately in light of new empirical evidence. In other words, Boyd argues that the underdetermination of astrophysical models is more often transient than permanent, and that the epistemic status of such models therefore remains significant.

Crowther, K., Linnemann, N. S., & Wüthrich, C. (2021). What we cannot learn from analogue experiments. *Synthese, 198*(S16), 3701–3726. https://doi.org/10.1007/s11229-019-02190-0.

Contrary to Dardashti et al. (2017; 2019), Thébault (2019), and Evans and Thébault (2020), Crowther et al. argue that analogue experiments used to investigate inaccessible target phenomena (for instance, fluid "dumb holes" used to investigate astrophysical black holes) are no more confirmatory than analogical arguments – which is to say, hardly confirmatory at all. More specifically, the authors argue that analogue experiments cannot confirm whether a particular inaccessible phenomenon (such as Hawking radiation) actually exists, and they criticize their opponents for unjustifiably assuming the physical adequacy of analogue modelling frameworks, thereby begging the question. Despite this, the authors admit that analogue experiments can be useful scientific tools for exploring the relevant modeling framework and for demonstrating robustness of the phenomena of which they are designed to be analogues.

Dardashti, R., Thébault, K. P. Y., & Winsberg, E. (2017). Confirmation via analogue simulation: What dumb holes could tell us about gravity. *The British Journal for the Philosophy of Science, 68*(1), 55–89. https://doi.org/10.1093/bjps/axv010.

Using Hawking radiation as a case study, the authors argue that analogue models of inaccessible astrophysical phenomena can be used to confirm predictions about such phenomena given (1) a robust syntactic isomorphism between the modelling frameworks of the analogue and the target systems, (2) diverse analogue realizations of the phenomenon under study, and (3) valid universality arguments.

Dardashti, R., Hartmann, S., Thébault, K. P. Y., & Winsberg, E. (2019). Hawking radiation and analogue experiments: A Bayesian analysis. *Studies in History and Philosophy of Science Part B: Studies in History and Philosophy of Modern Physics, 67,* 1–11. https://doi.org/10.1016/j.shpsb.2019.04.004

Extending Dardashti et al. (2017)'s discussion of universality arguments, the authors provide a quantitative Bayesian model for investigating the inferential structure and confirmatory power of analogue black hole experiments. Their formal model shows how to link evidence about analogue systems to target systems, accounts for the confirmatory relevance of "saturation" (when multiple types of analogues are used to probe the same targets), and shows that the more confident we are about the physics underlying a particular analogue, the less it can teach us about the target system.

"Hawking process"), to explore the intrinsically interesting behavior of the analogue systems themselves, and to investigate the robustness of predicted phenomena which may contribute to a two-way knowledge flow between analogue and target systems. The paper also contains a helpful discussion of the history of analogue black hole experiments, and explains how old experiments may be reinterpreted using the author's framework

Gueguen, M. (2020). On robustness in cosmological simulations. *Philosophy of Science, 87*(5), 1197–1208. https://doi.org/10.1086/710839

Scientists use numerical simulations to determine the mass distribution of dark matter halos. Numerical results are normally taken to be confirmatory when they are robust, i.e., when they resist some degree of fluctuation in the values of certain underlying parameters, as typically explored in "convergence studies". However, Gueguen (Institute of Physics of Rennes 1) argues that robustness analysis in the form of convergence studies fails to exclude numerical artifacts, and that in fact convergence can *result* from artifacts; we need a better criterion for determining the trustworthiness of our simulations.

Gueguen, M. (2021). A tension within code comparisons [unpublished manuscript]. http://philsci-archive.pitt.edu/19227/.

While convergence studies like those discussed in Gueguen (2020) are meant to test for the "internal robustness" of astrophysical simulations, code comparisons (which look for shared results across different simulation codes) appear to test for "external robustness". However, Gueguen (Institute of Physics of Rennes 1) argues that the presence of shared results across different astrophysical simulations has little epistemic significance in practice, and that even in principle (with a perfectly constructed ensemble of codes to compare), the requirement that the codes bear on comparable targets is inevitably in tension with the requirement that they differ with respect to their components. Thus, code comparisons cannot help us decide whether to trust a given simulation.

Jacquart, M. (2020). Observations, simulations, and reasoning in astrophysics. *Philosophy of Science, 87*(5), 1209–1220. https://doi.org/10.1086/710544.

Using collisional ring galaxies as a case study, Jacquart (University of Cincinnati) argues that computer simulations in astrophysics play three epistemic roles: (1) hypothesis testing (eg., testing possible explanations for how a galaxy could form a ring-shape), (2) exploring possibility space (eg., to establish the parameter boundaries in which ring galaxy-formation occurs), and (3) amplifying observations (i.e., using the simulation to develop a context in which to interpret observational data).

Jebeile, J. (2017). Computer simulation, experiment, and novelty. *International Studies in the Philosophy of Science, 31*(4), 379-395. https://doi.org/10.1080/02698595.2019.1565205.

Can computer simulations provide genuinely new knowledge? Using the "dark" galaxy called VirgoHI21 as a test case, Jebeile (University of Bern) argues affirmatively that although only concrete experiments can confound scientists and refute theories, simulations can still provide new knowledge *qua* knowledge obtained for the first time which adds to existing knowledge (this is the "first time" criterion of novelty). Importantly, the ability of simulations to generate new knowledge does *not* depend on features that they share with experiments.

Jebeile, J., & Kennedy, A. G. (2015). Explaining with models: The role of idealizations. *International Studies in the Philosophy of Science, 29*(4), 383–392. https://doi.org/10.1080/02698595.2015.1195143.

On the typical representationalist view of model explanation, idealized models are less explanatory than de-idealized models. Using galactic simulations as a case study, Jebeile (University of Bern) and Kennedy (Florida Atlantic University) contend that de-idealization is not always *in itself* explanatorily beneficial; sometimes, comparisons between idealized and de-idealized models allow researchers to extract important explanatory information not available in the de-idealized model alone. Furthermore, the authors argue that model explanation ought to be understood not as a product or feature of models, but as a user-dependent *activity*.

Massimi, M. (2018a). Perspectival modeling. *Philosophy of Science, 85*, 335-359. https://doi.org/10.1086/697745.

It is intuitive that using a plurality of models to represent one target system stifles the quest for scientific realism. However, Massimi (University of Edinburgh) argues that the problem of inconsistent models can be solved if we reconceptualize the role of models as representing not actual or fictional states of affairs, but *possibilities* in a possibility space. If we do so, using and testing a plurality of models can help narrow this space, which is inherently valuable for the realist goal of achieving true or approximately true theories. This article provides an interesting rebuttal to the model anti-realism of Ruphy (2011), and represents a potentially fruitful framework for interpreting inconsistent astrophysical models.

Meskhidze, H. (2017). Simulationist's regress in laboratory astrophysics [unpublished manuscript].

Extending the idea of the "experimenter's regress" from Collins (1985) and of the "simulationist's regress" from Gelfert (2011),[1] Meskhidze (UC Irvine) argues that the widespread use of modular models and bootstrapping methods in astrophysics renders the field susceptible to irresolvable situations of regress. The paper also includes interesting discussions of internal, external, and construct validity.

---

[1] Gelfert, A. (2011). Scientific models, simulation, and the experimenter's regress. In P. Humphreys & C. Imbert (Eds.), *Models, Simulations, and Representations* (pp. 145-167). Routledge.

Reutlinger, A., Hangleiter, D., and Hartmann, S. (2018). Understanding (with) toy models. *The British Journal for the Philosophy of Science, 69*(4), 1069-1099. https://www.doi.org/10.1093/bjps/axx005.

Can simplified and idealized scientific models ("toy models") provide genuine understanding? In this paper, the authors divide such models into two types – those which are *embedded* within an empirically well-confirmed framework theory, and those which are *autonomous* from any such framework – and argue that the former can provide "how-actually" understanding and the latter "how-possibly" understanding. Given that astrophysical models are sometimes quite simplified and idealized, this article provides a framework for understanding the epistemic role of such models.

Ruphy, S. (2011). Limits to modeling: Balancing ambition and outcome in astrophysics and cosmology. *Simulation & Gaming: An Interdisciplinary Journal, 42,* 177-194. https://doi.org/10.1177/1046878108319640.

Ruphy (École normale supérieure – PSL) argues that in galactic astrophysics, there are often numerous empirically adequate submodels available for researchers to choose from, and the choice of a particular submodel at a given stage constrains the range of available submodels at later stages. This renders models path-dependent and contingent. Combined with the plasticity and stability of such models, these features can lead to persistent incompatible model pluralism, which thwarts the goal of accurately representing the world; accordingly, we should be anti-realists about galactic models.

Thébault, K. (2019). What can we learn from analogue experiments? In R. Dardashti, R. Dawid, and K. Thébault (Eds.), *Why Trust a Theory? Epistemology of Fundamental Physics* (pp. 184-201). Cambridge University Press. https://doi.org/10.1017/9781108671224.014.

Analogue experiments, for example the use of fluid models to investigate Hawking radiation, can provide us with evidence of the same confirmatory type (and plausibly even of the same confirmatory degree) as conventional experiments. This is because, according to Thébault (University of Bristol) we can externally validate analogue black holes and thus take them to stand in for their astrophysical cousins.

*For further articles relevant to this category, see Anderl 2016, Elder 2020, Elder 2021/2, Elder 2022, Meskhidze 2021, Salmon 1998, Suárez 2013, Sundberg 2010, Sundberg 2012, and Wilson 2021.*

**3. Realism and antirealism**

much of a science as any other, and that Hacking's antirealism depends on an overly static understanding of science. Furthermore, Shapere argues that Hacking's 1989 article on gravitational lenses cherry picks its data, interprets these data too narrowly, and falsely concludes that the use of incompatible models renders realistic treatment of astrophysical phenomena impossible.

Suárez, M. (2013). Fictions, conditionals, and stellar astrophysics. *International Studies in the Philosophy of Science, 27*(3), 235–252. https://doi.org/10.1080/02698595.2013.825499.

Using models of stellar structure as a case study, Suárez (Complutense University of Madrid) contends that the main assumptions of such models are best understood as useful fictions, but that scientists can nevertheless maintain a realist agenda by (1) treating such assumptions as background knowledge required for the generation of "fictional conditionals", or (2) treating such assumptions as components of the antecedents of these conditionals and employing a non-truth-functional semantics for them.

*For further articles relevant to this category, see Anderl 2016, Boyd 2015, Hudson 2007, Massimi 2018, and Ruphy 2011.*

## 4. Theories and testing

Boyd, N. M. (2018). *Scientific Progress at the Boundaries of Experience.* Ph.D. dissertation. University of Pittsburgh. http://d-scholarship.pitt.edu/id/eprint/33843.

In this PhD dissertation, Boyd (Siena College) articulates a new empiricist philosophy of science and a non-internalist conception of scientific progress according to which the accumulating corpus of empirical data available to us (despite its theory-ladenness) constrains viable theories and constitutes growing knowledge about the world. Although Boyd offers a general account of scientific progress, her discussion is largely furnished with examples from the observational sciences, especially astrophysics and cosmology. Case studies include Arecibo telescope data, Babylonian astronomical tables, dark energy, and cosmic inflation. For the published version of the dissertation's third chapter, see Boyd, N. (2018). Evidence enriched. *Philosophy of Science 85*, 403-421. https://doi.org/10.1086/697747.

De Baerdemaeker, S., & Boyd, N. M. (2020). Jump ship, shift gears, or just keep on chugging: Assessing the responses to tensions between theory and evidence in contemporary cosmology. *Studies in History and Philosophy of Science Part B: Studies in History and Philosophy of Modern Physics, 72,* 205–216. https://doi.org/10.1016/j.shpsb.2020.08.002

When comparing predictions from the ΛCDM model with high-resolution astronomical observations, we face three dark matter-related "small-scale challenges": the Missing

Satellites problem, the Too Big to Fail problem, and the Cusp/Core problem. De Baerdemaeker (Stockholm University) and Boyd (Siena College) note three potential responses scientists can take to these problems, namely to jump ship (i.e., abandon ΛCDM for something like MOND), to switch gears (i.e., modify ΛCDM with something like warm dark matter), or to keep on chugging (i.e., focus on improving ΛCDM simulations by incorporating known baryonic physics). Based on the heuristics of epistemic conservatism and individuating causal factors, the authors argue that scientists ought to keep on chugging, and they conclude by outlining potential future scenarios in dark matter research.

Principle) is common to all metric theories of gravity. Given that all viable theories of gravity are metric, any such theory can be employed when investigating dark matter lenses -- the specifics of GR or any other theory are only required to determine the amount of dark matter present. Thus, contrary to the "dark matter double-bind" proposed by Vanderburgh (2003; 2005), Kosso claims that dark matter can be detected without assuming the truth of GR. See Sus (2014) and Vanderburgh (2014b) for responses.

Martens, N. C. M., & Lehmkuhl, D. (2020a). Dark matter = modified gravity? Scrutinising the spacetime–matter distinction through the modified gravity/dark matter lens. *Studies in History and Philosophy of Science Part B: Studies in History and Philosophy of Modern Physics, 72*, 237–250. https://doi.org/10.1016/j.shpsb.2020.08.003.

Most proposed solutions to the dark matter problem are either matter-based (eg., WIMPS) or gravity-based (eg., MOND). These types of solutions are typically represented as conceptually distinct, owing to a deeper distinction between matter and spacetime. In this paper, Martens and Lehmkuhl (both University of Bonn) argue that a strict matter-spacetime distinction is untenable, and likewise for the distinction between matter vs. gravity-based solutions to the dark matter problem. Their analysis draws heavily from the recent literature on superfluid dark matter, the scalar field φ of which they interpret both as a kind of dark matter, and as a modification of gravity. This paper constitutes the first part of a pair of articles, the second being Martens and Lehmkuhl (2020b).

Martens, N. C. M., & Lehmkuhl, D. (2020b). Cartography of the space of theories: An interpretational chart for fields that are both (dark) matter and spacetime. *Studies in History and Philosophy of Science Part B: Studies in History and Philosophy of Modern Physics, 72*, 217–236. https://doi.org/10.1016/j.shpsb.2020.08.004.

Following up from Martens and Lehmkuhl (2020a), the authors advance a "cartographic" taxonomy of interpretations for "Janus-faced" theories like superfluid dark matter (according to which a single scalar field is both a dark matter field and a modification of gravity, in certain contexts). Their taxonomy contains three classes of interpretations with nine subclasses, and they argue that four such subclasses remain viable ways of understanding superfluid dark matter. See p. 231 for their chart of interpretations.

Matarese, V. (2022). 'Oumuamua and meta-empirical confirmation. *Foundations of Physics, 52*(4). https://doi.org/10.1007/s10701-022-00587-5.

Astrophysicist Abraham Loeb has suggested that the interstellar interloper 1I/2017 'Oumuamua is a piece of alien technology. To empirically confirm or confute his hypothesis would require significant expenditure of financial and intellectual resources – for instance by sending a probe to 'Oumuamua, as proposed by *Project Lyra.* How can we be sure that Loeb's hypothesis is viable and thus worth pursuing at all? Matarese (University of Bern) argues that we should use a meta-empirical framework to answer this question, one which provides information about the capacity of Loeb's hypothesis to adequately represent potential future empirical data. Furthermore, Matarese contends that meta-empirical confirmation does not violate the empiricist spirit since it can be fruitfully applied even in empirically grounded research contexts such as this one.

attempts to identify and evaluate patterns of inference involved in evidential arguments for candidate solutions thereof.

---

[2] Baker, A. (2003). Quantitative parsimony and explanatory power. *British Journal for the Philosophy of Science, 54*, 245-259. https://doi.org/10.1093/bjps/54.2.245.

> Peter Kosso (2013) has argued that evidence from observations of the Bullet Cluster provides evidential warrant for the equivalence principle in such a way which avoids Vanderburgh's (2001; 2003) "dark matter double bind". Vanderburgh (CSU San Bernardino) responds that even if this is the case, we are still unable to perform the kind of precision tests of general relativity that would confirm its applicability to galactic and supra-galactic scales. Vanderburgh also countenances the possibility that we cannot rule out "ugly" solutions to the dark matter problem which incorporate both dark matter and modified theories of gravity.

## 5. SSK and social issues

> A classic text in the sociology of scientific knowledge which introduces the notion of the "experimenter's regress" using Joseph Weber's attempts to detect gravitational waves as a case study. According to Collins (Cardiff University), it often happens in frontier science that the best or only check on a result is the proper functioning of the apparatus used to generate it, while the best of only check on the proper functioning of the apparatus is the result; this is the experimenter's regress. For Collins, there is usually no rational way out of the regress – instead, scientists resort to heuristics, rhetoric, compulsion, etc. For those interested specifically in the Weber case study and regress, see Chapter 4 (pp. 79-111).

> The authoritative social history of gravitational waves from the 1960s-2004, Collins' (Cardiff University) book touches on key issues of scientific knowledge, expertise, and consensus. It serves as an important resource for philosophers interested in gravitational waves who seek to understand the scientific process at a more concrete level. For Collins' other books on the science and sociology of gravitational waves, see Collins, H. (2010). *Gravity's Ghost: Scientific Discovery in the Twenty-first Century*. University of Chicago Press; and Collins, H. (2017). *Gravity's Kiss: The Detection of Gravitational Waves*. MIT Press.

epistemic conflict between the disciplines, leading to transgression of social and cognitive boundaries, turbulence in epistemic practices, and self-reflection on the scientists' identities and the moral economy of which they are a part.

Jebeile, J. (2018). Collaborative practice, epistemic dependence and opacity: The case of space telescope data processing. *Philosophia Scientiæ. Travaux d'histoire et de Philosophie Des Sciences, 22*(2), 59-78. https://doi.org/10.4000/philosophiascientiae.1483.

Employing Susann Wagenknecht's (2014) distinction between opaque and translucent epistemic dependence,[3] Jebeile (University of Bern) analyzes the social epistemological relationships between collaborators working with the Herschel Space Observatory. Jebeile identifies cases of opaque epistemic dependence therein, and argues that sources of opacity include not only lack of expertise, but also the non-disclosure of data, failure to understand relevant instrumental processes, and epistemic inaccessibility of numerical calculations.

Kennefick, D. (2000). Star crushing: Theoretical practice and the theoretician's regress. *Social Studies of Science, 30*(1), 5-40. https://doi.org/10.1177/030631200030001001.

An important study in the sociology of astrophysical knowledge and simulations which uses the 1990s controversy over "star-crushing" as a case study. Kennefick (University of Arkansas) introduces the notion of the "theoretician's regress" – a play on Harry Collins' "experimenter's regress" (Collins 1985) – to explain why the controversy over the star-crushing effect could not be resolved by strictly "scientific" debate.

McCray, W. P. (2000). Large telescopes and the moral economy of recent astronomy. *Social Studies of Science, 30*(5), 685-711. https://doi.org/10.1177/030631200030005002.

McCray (UC Santa Barbara) contends that the Gemini 8-Meter Telescopes Project illustrates a tension in American optical astronomy between the "haves" (those with access to telescopes through their institutions) and the "have-nots" (those who must compete for time at federally funded national observatories), while also showing how non-scientific political concerns play into the debate around investment in big science.

Messeri, L. R. (2010). The problem with Pluto: Conflicting cosmologies and the classification of planets. *Social Studies of Science, 40*(2), 187-214. https://doi.org/10.1177/0306312709347809.

---

[3] Wegenknecht, S. (2014). Opaque and translucent epistemic dependence in collaborative scientific practice. *Episteme, 11*(4), 475-492. https://www.doi.org/10.1017/epi.2014.25. According to Wegenknecht:
"A scientist is opaquely dependent upon a colleague's labor, if she does not process the expertise necessary to independently carry out, and to profoundly assess, the piece of scientific labor her colleague is contributing. I suggest, however, that if the scientist does possess the necessary expertise, then her dependence would not be opaque, but translucent". (p. 483)

This paper offers a historical and sociological account of the "demotion" of Pluto by the IAU in 2006. Messeri (Yale University) focuses on the relationships between scientific and cultural cosmologies, and the ways that the public influenced the debate surrounding the definition of "planet". Messeri argues that the IAUs definition of "planet" privileged one cosmology over others, thereby fracturing discourse about planets and Pluto in particular.

Metzger, P. T., Grundy, W. M., Sykes, M. V., Stern, A., Bell, J. F., Detelich, C. E., Runyon, K., & Summers, M. (2022). Moons are planets: Scientific usefulness versus cultural teleology in the taxonomy of planetary science. *Icarus, 374*, 114768. https://doi.org/10.1016/j.icarus.2021.114768.

In this lengthy and detailed article, Metzger et al. contend that the IAUs 2006 definition of "planet" paid too much credence to folk taxonomy while ignoring the importance of scientific taxonomy. They argue that a purely geophysical definition of "planet", according to which a planet is an object with a certain amount of geological complexity, has stronger historical and pragmatic grounding when compared to the current dynamical definition (and that contemporary planetary scientists already use the geophysical definition anyways). As such, the authors claim that we ought to revise our educational materials for the sake of an improved scientific and cultural planetary taxonomy.

Sovacool, B. (2005). Falsification and demarcation in astronomy and cosmology. *Bulletin of Science, Technology & Society, 25*(1), 53–62. https://doi.org/10.1177/0270467604270151.

In this sociological article, Sovacool (University of Sussex; Aarhus University) analyzes how and to what extent contemporary astronomers and cosmologists rely on the ideas of Karl Popper to resolve crises related to methodology, legitimacy, and testability. Sovacool concludes that Popper's ideas play an important implicit and explicit role in such crises, and he argues more broadly for the relevance of philosophy to scientific practice.

Sundberg, M. (2010). Cultures of simulations vs. cultures of calculations? The development of simulation practices in meteorology and astrophysics. *Studies in History and Philosophy of Science Part B: Studies in History and Philosophy of Modern Physics, 41*, 273-281. https://doi.org/10.1016/j.shpsb.2010.07.004.

Sundberg (Stockholm University) uses "modern" and "postmodern" "computer cultures" as a lens for analyzing contemporary numerical simulation practices in meteorology and astrophysics. Sundberg argues that, by and large, there seems to be a shift occurring towards a more "postmodern" culture of simulations which emphasizes the value of surface-level research using black-box simulations, entertaining visualizations of simulation results, and the playful exploration of simulation codes as a learning tool.

Sundberg, M. (2012). Creating convincing simulations in astrophysics. *Science, Technology, & Human Values, 37*(1), 64–87. https://www.doi.org/10.1177/0162243910385417.

A study in the sociology of scientific knowledge, Sundberg (Stockholm University) examines the methods that astrophysicists use to convince themselves and others of the reliability and credibility of their simulations, especially when those simulations deliver outputs which are uncertain or difficult to interpret. In the process, Sundberg analyzes the distinction between "numerical" and "real" effects, arguing that they cannot be distinguished on the basis of what they derive from.

*For another article relevant to this category, see Hudson 2007.*

## 6. Typicality and extra-terrestrials

Ćirković, M. M. (2006). Too early? On the apparent conflict of astrobiology and cosmology. *Biology and Philosophy, 21*(3), 369–379. https://doi.org/10.1007/s10539-005-8305-2.

Olum's problem, a generalized form of Fermi's paradox, relies on a dehistoricized understanding of the universe. In opposition to this dehistoricized stance, Ćirković (Astronomical Observatory of Belgrade; Future of Humanity Institute) argues that the universe becomes more hospitable to life as time passes, meaning that it is *too early* in cosmological history for the absence of detected extraterrestrial life to constitute a genuine paradox.

Lacki, B. (2021). The noonday argument: Fine-graining, indexicals, and the nature of Copernican reasoning [unpublished manuscript]. arXiv:2106.07738v1 [physics.hist-ph].

Lacki (UC Berkeley) offers a new theory of typicality called Fine Graining with Auxiliary Indexicals (FGAI). He argues that it avoids the paradoxes (such as the Doomsday Argument) faced by other theories of typicality by fine-graining our macrotheories with microhypotheses and by separating indexical from physical facts.

Lewis, G. F. & Barnes, L. A. (2021). The trouble with "puddle thinking:" A user's guide to the Anthropic Principle. *Journal & Proceedings of the Royal Society of New South Wales, 154*(1), 6-11. ISSN 0035-9173/21/010006-06.

A short, popular introduction to the anthropic principle and to Douglas Adams' notion of "puddle thinking", Lewis (University of Sydney) and Barnes (Western Sydney University) argue that the problem of fine-tuning does not concern the question of how we find ourselves in a universe with conditions amenable to human life, but the question of *why* a universe with such conditions exists at all. This question is, they concede, a necessarily philosophical one.

Satta, M. (2021). Evil twins and the multiverse: Distinguishing the world of difference between epistemic and physical possibility. *Synthese, 198*(2), 1153–1160. https://doi.org/10.1007/s11229-019-02092-1.

Brian Greene and Max Tegmark have claimed that if the universe is infinite and matter is roughly evenly distributed within it, then every possible material arrangement of particles must exist in an infinite number of instantiations. Satta (Wayne State University) argues that Green and Tegmark's claims rely on a conflation of physical with epistemic possibility, and that they ignore potential macro-level constraints on possibility from psychology, biology, and sociology.

*For another article relevant to this category, see Matarese 2022.*

## 7. Dark matter and MOND

Abelson, S.S. (2022). The fate of tensor-vector-scalar modified gravity. *Foundations of Physics, 52*(31). https://doi.org/10.1007/s10701-022-00545-1.

Abelson (Indiana University, Bloomington) reviews the case of the LIGO neutron star merger detection against TeVeS, long considered the most plausible relativistic extension of MOND. Abelson argues that the physicists' use of language of falsification in a strict Popperian sense was unwarranted. However, Abelson offers an alternative interpretation of the result as a corroboration of the null-hypothesis, along the lines of Mayo's error-statistical account.[4]

De Baerdemaeker, S. & Dawid, R. (2022) MOND and meta-empirical theory assessment. *Synthese 200*(344). https://doi.org/10.1007/s11229-022-03830-8

De Baerdemaeker and Dawid (both Stockholm University) critically examine some of the philosophical arguments that have been offered by defenders of MOND in terms of different views on theory assessment. They argue, first, that on a standard reading of Popper and Lakatos, the arguments fail. Second, they argue that the strongest philosophical defense of MOND takes the form of meta-empirical theory assessment, but that, according to that account as well, the arguments fail to be convincing.

Jacquart, M. (2021). ΛCDM and MOND: A debate about models or theory? *Studies in History*

---

[4] Mayo, D. (2018). *Statistical Inference as Severe Testing: How to Get Beyond the Statistics Wars*. Cambridge University Press. https://doi.org/10.1017/9781107286184.

McGaugh (Case Western University) reviews the ongoing disagreement between ΛCDM and MOND in terms of a Kuhnian picture of incommensurable paradigms. McGaugh submits that both have significant empirical support, but nonetheless offer inconsistent worldviews. However, McGaugh is concerned about the detectability of dark matter: without a positive dark matter particle detection, there are concerning parallels between the dark matter hypothesis and aether theory.

Merritt, D. (2017). Cosmology and convention. *Studies in History and Philosophy of Science Part B: Studies in History and Philosophy of Modern Physics, 57*, 41-52. https://doi.org/10.1016/j.shpsb.2016.12.002.

Merritt (Rochester Institute of Technology) assesses the current concordance model of cosmology according to Popper's definition of conventionalist stratagems. According to Merritt, dark matter and dark energy were ad hoc auxiliary hypotheses introduced to save the concordance model from falsifying evidence. Moreover, the usual convergence arguments offered in support of the concordance model are argued to fail. As such, cosmology has, plausibly, entered the phase of what Lakatos called a degenerative problem shift.

Merritt, D. (2020). *A Philosophical Approach to MOND: Assessing the Milgromian Research Program in Cosmology*. Cambridge University Press, Cambridge. https://doi.org/10.1017/9781108610926

Merritt (Rochester Institute of Technology) offers a book-length review of the history of MOND, from when it was first proposed in the early 1980s, until today. The philosophical framing of the book is in terms of Lakatos' theory of progressive and degenerative research programs. Merritt argues that especially the earlier theories of the research program were highly successful at predicting novel facts (assessed according to different proposed philosophical accounts of novelty). While the very latest theories are potentially less successful qua novel prediction, the overall research program is deemed to be progressive.

Merritt, D. (2021) Feyerabend's rule and dark matter. *Synthese* 199, 8921–8942. https://doi.org/10.1007/s11229-021-03188-3

Building on Feyerabend's work, Merritt's (Rochester Institute of Technology) starting point is the claim that, under specific circumstances, the lack of an experimental results can refute a theory while confirming another. Merritt applies this to the current concordance model of cosmology, and argues that there are several examples of such refuting negative results, including the failure of dark matter particle detection, the failure to detect primordial dwarf galaxies, and dynamical friction.

Milgrom (Weizmann Institute) reviews the case for MOND, from its initial proposal, up until its scientific status today. The paper draws multiple comparisons between the history of MOND and well-known episodes in the history of science, including the Copernican revolution and the development of quantum theory. It also draws parallels between dark matter and aether-theory.

*For further articles relevant to this category, see De Baerdemaeker 2021, De Baerdemaeker and Boyd 2020, Horvath 2009, Hudson 2007, 2009, 2013, Kosso 2013, Martens 2022, Martens and Lehmkuhl 2020a,b, Sus 2014, and Vanderburgh 2001, 2003, 2005, 2014a,b, Wilson 2021.*